\documentclass[epj]{svjour}

\usepackage{graphicx}
\usepackage{fancyhdr}
\def\makeheadbox{{
 }}
\def\mytitle{My title} 
\def\myauthors{My name}  
\def\mytype{My type of session}
\def\mysession{My session}

\def\mytitle{SUSY Parameters Determination with ATLAS} 
\def\myauthors{Nurcan \"Ozt\"urk}    
\def\mytype{Contributed Talk}    
\def\mysession{Colliders - SUSY Phenomenology}

\begin{document}
\title{SUSY Parameters Determination with ATLAS}
\author{Nurcan \"Ozt\"urk for the ATLAS Collaboration
}                     
%
%
\institute{Department of Physics, University of Texas at Arlington, Arlington, TX, 76019, USA
%
}
%
\date{}
\abstract{
The plan for mass and spin measurement of SUSY particles with the ATLAS detector is presented. 
The measurements of kinematical distributions, such as edges in the invariant mass of leptons 
and jets, could be used to constrain the model of SUSY that may be discovered at the LHC. 
Examples from a few points in the mSUGRA scenario are provided with an emphasis on measurements 
that can be conducted within the first few years of data taking.
} 
\maketitle
\section{Introduction}
Discovering Supersymmetry (SUSY) is one of the motivations for building the Large Hadron Collider 
(LHC) which is scheduled to take data in 2008. ATLAS (A Torodial LHC ApparatuS) is one of the two 
general purpose experiments at the LHC. Currently the ATLAS experiment is revisiting its SUSY 
studies in the framework of the CSC (``Computing System Commissioning''). For this exercise event 
samples for a set of SUSY benchmark points were produced with the detailed simulation of the ATLAS 
detector. The aim of the work is to better understand the impact of realistic experimental conditions 
including trigger efficiencies and imperfect calibration/alignment on the SUSY potential of the experiment.     

If SUSY is discovered by inclusive searches, next step will be to measure the masses and the spins of 
the SUSY particles through the analysis of exclusive decay channels. In this note SUSY mass and spin 
measurement techniques and methods for determining underlying SUSY model parameters are described within 
the mSUGRA framework.

\section{mSUGRA Framework}
The minimal SUSY extension of the SM (MSSM) brings 105 additional free parameters into 
the theory thus making a systematic study of the full parameter space difficult. A specific 
well-motivated model framework is usually assumed in which generic signatures can be studied. 
In the minimal Supergravity (mSUGRA) framework, SUSY is broken by the gravitational interactions 
and the masses and couplings are unified at the grand unified energy scale giving five free 
parameters; $m_0$ and $m _{1/2}$ (universial scalar and gaugino mass parameters), 
$A_0$ (universial trilinear coupling), ${\rm tan}(\beta)$ (the ratio of the vacuum 
expectation values of the scalar fields) and ${\rm sgn}(\mu)$ (the sign of the higgsino mass term). 
In the R-parity conserving mSUGRA models, SUSY particles are produced in pairs and the lightest 
SUSY particle (LSP) is stable and undetected resulting in large missing transverse energy 
(dominant signature). It will be possible to detect SUSY in multiple inclusive signatures over 
a large part of the mSUGRA parameters space with just 1 fb$^{-1}$ of integrated luminosity, as 
discussed in~\cite{Yamamoto}. As a part of the ongoing CSC exercise, several mSUGRA points 
that are favored by the WMAP data~\cite{WMAP} have been simulated with a detailed detector 
simulation. The mSUGRA parameters for these points are given in Table~\ref{tab:mSUGRA}.
\begin{table*}
\caption{mSUGRA points chosen for full detector simulation in the CSC exercise. 
ISAJET 7.71 is used and the top quark mass is set to 175 GeV.}
\label{tab:mSUGRA}      
\begin{center}
\begin{tabular}{llllllll}
\hline\noalign{\smallskip}Point Name & mSUGRA Region & $m_0$ (GeV) & $m_{1/2}$ (GeV) & $A_0$ (GeV) & tan($\beta$) & sgn($\mu$) & $\sigma$ (pb)   \\
\noalign{\smallskip}\hline\noalign{\smallskip}
SU1 & Coannihilation  & 70 & 350 & 0 & 10 & + & 7.43\\
SU2 & Focus & 3550 & 300 & 0 & 10 & + & 4.86 \\
SU3 & Bulk & 100 & 300 & -300 & 6 & + & 18.59 \\
SU4 & Low mass & 200 & 160 & -400 & 10 & + & 262 \\
SU6 & Funnel & 320 & 375 & 0 & 50 & + & 4.48 \\
SU8.1 & Coannihilation & 210 & 360 & 0 & 40 & + & 6.44 \\
SU8.2 & Coannihilation & 215 & 360 & 0 & 40 & + & 6.40 \\
SU8.3 & Coannihilation & 225 & 360 & 0 & 40 & + & 6.32 \\
\noalign{\smallskip}\hline
\end{tabular}
\vspace*{-0.5cm}  
\end{center}
\end{table*}
\section{Mass Measurement Techniques}
\label{sec:2}
In the R-parity conserving mSUGRA models all SUSY events contain two invisible neutralinos
(${\widetilde \chi}_{1}^{0}$) which, being only weakly interacting, escape the detector; 
therefore no mass peaks can be reconstructed directly. However kinematic endpoints and 
thresholds in the invariant mass distributions of the visible decay products can be measured. 
The values of these kinematic features can be expressed as a function of the masses of the 
involved sparticles~\cite{Allanach}. The mass measurement strategy is to exploit kinematics of 
long decay chains originating from gluino or squark production. Specifically, from the 
reconstruction of the decay chain ${\widetilde q}_{L} \rightarrow q {\widetilde \chi}_{2}^{0} 
\rightarrow q {\widetilde l}_{R}^{\pm}l^{\mp} \rightarrow q {\widetilde \chi}_{1}^{0} 
l^{\pm} l^{\mp}$ 
one can measure the masses of ${\widetilde q}_{L}$, ${\widetilde l}_{R}$, 
${\widetilde \chi}_{2}^{0}$ and ${\widetilde \chi}_{1}^{0}$. Examples 
of mass measurements using the endpoint method are given below.

\subsection{Dilepton Endpoint}
\begin{figure}
\begin{center}
\includegraphics[width=0.41\textwidth,height=0.28\textwidth,angle=0]{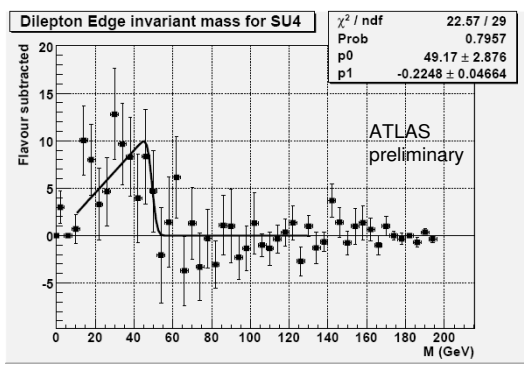}
\caption{The dilepton invariant mass distribution for a full simulation sample of the 
SU4 point with an integrated luminosity of 0.35 fb$^{-1}$. A triangular function convoluted 
with a Gaussian is fitted to estimate the edge position.}
\label{fig:su4_dilepton}    
\end{center}
\end{figure}

A clear endpoint is exhibited by the dilepton invariant mass distribution for the decay channel 
${\widetilde \chi}_{2}^{0}\rightarrow {\widetilde l}_{R}^{\pm}l^{\mp} \rightarrow{\widetilde \chi}_{1}^{0}
l^{\pm} l^{\mp}$. Fully simulated samples for the SU4 point (see Table~\ref{tab:mSUGRA}) 
and $t\bar{t}$ background are considered. Events are selected by requiring at least two 
leptons ($e$ or $\mu$) with $p_{T}>$10 GeV and $|\eta|<$2.5, a calorimetric energy deposit 
$E_{T}<$5 GeV in a $\eta-\phi$ cone of size 0.3 around the lepton direction, and at least 
four jets with $E_{T}>$100, 50, 50, 50 GeV and the missing transverse energy $E_{T}^{\rm miss}>$120 GeV. 
The effective mass variable $M_{\rm eff}$, defined as 
$M_{\rm eff}=E_{T}^{\rm miss} + \sum_{\rm jet} E_{T,\rm jet}$ 
taking into account the first four leading jets, is required to be $M_{\rm eff}>$550 GeV. 
The combinatorial background from $t\bar{t}$ and ${\widetilde \chi}^{\pm}$ decays cancel in the 
combination $e^{+}e^{-}+\mu^{+}\mu^{-}-e^{\pm}\mu^{\mp}$ (so called flavor subtraction) which is
plotted in Figure~\ref{fig:su4_dilepton}. The dilepton invariant mass distribution is shown for 
an integrated luminosity of 0.35 fb$^{-1}$. The mass distribution is fitted to a triangular function 
convoluted with a Gaussian to estimate the edge position. The endpoint fit gives a value of 
(49.2$\pm$2.9) GeV which is within 1.6$\sigma$ of the expected value of 53.7 GeV. The signal 
significance is calculated as 16.5 for 100 pb$^{-1}$ of data.

\subsection{Lepton-Jet Endpoint}
\begin{figure}
\begin{center}
\includegraphics[width=0.41\textwidth,height=0.29\textwidth,angle=0]{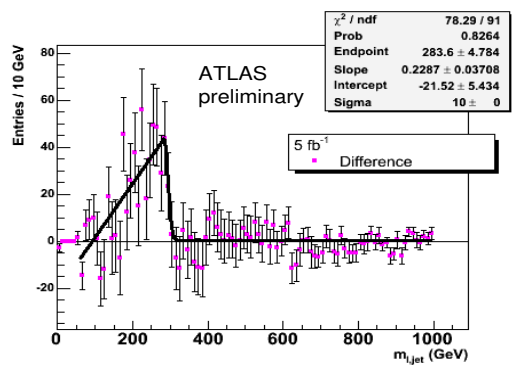}
\caption{The invariant mass distribution of lepton-jet for the SU1 full simulation sample 
with an integrated luminosity of 5 fb$^{-1}$. A triangular function convoluted with a 
Gaussian is fitted to estimate the edge position.}
\label{fig:su1_leptonjet}      
\end{center}
\end{figure}

For the SU1 point, the decay ${\widetilde q}_{L} \rightarrow q {\widetilde \chi_{1}^{\pm}\rightarrow 
q l^{\pm}{\widetilde \nu}_{l}} \rightarrow q l^{\pm}\nu_{l}{\widetilde \chi}_{1}^{0}$ 
gives rise to an endpoint in the lepton-jet invariant mass distribution. A mixed event technique 
is used to subtract the combinatorial jet background. The technique makes use of randomly pairing 
the jets from a different event (satisfying same event selection) with the lepton and then subtracting 
the mixed-event-jet distribution from the same-event-jets distribution (with a normalization 
correction applied) to obtain an inferred ``correct jet'' distribution. Fully simulated samples for 
the SU1 point and $t\bar{t}$ background are considered. Events are selected by requiring one lepton 
($e$ or $\mu$) with $p_{T}>$20 GeV and $|\eta|<$2.5, and a lepton isolation cut of $E_{T}<$10 GeV 
in a cone of 0.45. Additional cuts are applied to reduce the $t\bar{t}$ background; 
the leading and second leading jets are selected with $E_{T}>$200 GeV, the transverse mass 
$M_{T}<$60 GeV or $M_{T}>$100 GeV, $E_{T}^{\rm miss}>$250 GeV. 
The invariant mass of the lepton-jet is shown in Figure~\ref{fig:su1_leptonjet} for an integrated 
luminosity of 5 fb$^{-1}$. The mass distribution is fitted to a triangular function convoluted with a 
Gaussian to estimate the edge position. The endpoint fit gives a value of (283.6$\pm$4.8) 
GeV compared to the expected value of 284 GeV. The endpoint can be determined with 
5$\sigma$ statistical signal significance for 5 fb$^{-1}$ of data.

\subsection{Ditau Endpoint}
\begin{figure}
\begin{center}
\includegraphics[width=0.40\textwidth,height=0.28\textwidth,angle=0]{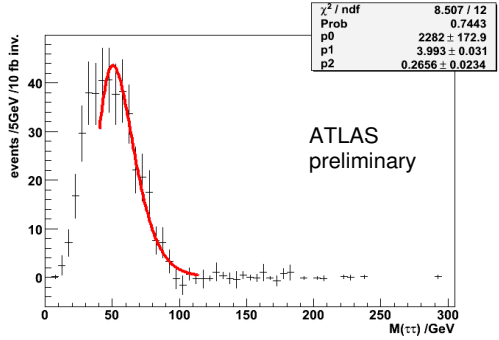}
\caption{The ditau invariant mass distribution for the SU3 fast simulation sample with 
an integrated luminosity of 10 fb$^{-1}$. It is fitted to a function as decribed in the 
text to estimate the endpoint.}
\label{fig:tau_1}
\end{center}     
\end{figure}

The mass difference between ${\widetilde \chi}_{2}^{0}$ and ${\widetilde \chi}_{1}^{0}$ can 
also be measured from the endpoint of the ditau invariant mass distribution from the decay 
channel ${\widetilde \chi}_{2}^{0}
\rightarrow {\widetilde \tau}_{1}^{\pm}\tau^{\mp} \rightarrow{\widetilde \chi}_{1}^{0}
\tau^{\pm} \tau^{\mp}$. The SU3 point has a factor of four larger branching ratio to ${\widetilde \chi}_{2}^{0}
\rightarrow{\widetilde \chi}_{1}^{0}\tau^{\pm} \tau^{\mp}$ than to $e/\mu$. 
A fast simulation sample for the SU3 point together with Z+jets, W+jets, 
$t\bar{t}$, $bb$+jets, dijets and multijets backgrounds is considered. Events are 
selected by requiring at least one jet with $E_{T}>$220 GeV, at least three jets with 
$E_{T}>$50 GeV and at least four jets with $E_{T}>$40 GeV and $E_{T}^{\rm miss}>$230 GeV. 
The angular seperation between the taus is chosen as $\Delta R(\tau,\tau)<$2. 
Only the hadronic tau decays are considered. The ditau invariant mass distribution 
calculated from $\tau^{+}\tau^{-}-\tau^{\pm}\tau^{\pm}$ (to cancel the background from 
${\widetilde \chi}_{1}^{\pm}$ decays) is shown in Figure~\ref{fig:tau_1} for an 
integrated luminosity of 10 fb$^{-1}$. Since the measurement of the endpoint by a linear fit 
has a strong dependence from the fitting range or binning of the distribution, a new approach is 
considered to estimate the endpoint. A fit function in the form of 
\begin{equation}
y=\frac{p_{0}}{x}.exp\left[\frac{-1}{2p_{2}^{2}}(ln(x)-p_{1})^{2}\right] 
\end{equation}
is used (modified from~\cite{CMS}). The inflection point is calculated as:
\begin{equation}
x_{IP}=exp\left[\frac{1}{2}p_{2}^{2}\left(-3+\sqrt{1+\frac{4}{p_{2}^{2}}}\;\right)+p_{1}\right]
\end{equation}
The measurement of the inflection point is more stable under variations of the fitting range or binning. 
A calibration line is driven between the inflection point and the endpoint by varying the 
involved masses in the decay chain. This calibration line is found to be $y=(0.47\pm0.02)x+(15\pm2)$. 
The endpoint value is then calculated as 
($105\pm4$) GeV compared to the expected value of 98.3 GeV. 
 
\section{mSUGRA Parameters Determination}
More endpoint measurements can be done for instance from the reconstruction of gluino decays;
${\widetilde g} \rightarrow {\widetilde \chi_{2}^0}t\bar{t}$ and ${\widetilde g} \rightarrow
{\widetilde t_{1}}t \rightarrow tb {\widetilde \chi_{1}^\pm}$ for which the details can be found
in~\cite{Sanctis} and~\cite{Krstic} respectively. Also, the mass of ${\widetilde q}_{R}$ can be
measured from the decay ${\widetilde q}_{R} \rightarrow {\widetilde \chi_{1}^0}q$ 
as studied in~\cite{Krstic}. The potential for extracting the masses from the endpoint measurements 
is described in~\cite{Borge} for the SPS1a point~\cite{Allanach:2002nj} with 300 fb$^{-1}$. 
The ultimate goal is to perform a fit of the mSUGRA parameters from a given set of mass measurements 
using the tools such as Fittino~\cite{FITTINO} and SFitter~\cite{SFITTER}. As an example, the measured masses 
and kinematical edges of the SPS1a point from~\cite{Borge} are fed into the SFitter program to determine 
the mSUGRA parameters for 300 fb$^{-1}$~\cite{Zerwas}. As seen in 
Table~\ref{tab:fit} the parameters can be determined with a precision at the percent level by 
using the masses ($\Delta_{masses}$ column), moreover the precision can be improved significantly 
by using the measured edges, thresholds and mass differences ($\Delta_{edges}$ column) in the fit 
instead of the masses.
\begin{table}
\caption{Determination of mSUGRA parameters for the SPS1a point using the SFitter program. 
${\rm sgn}(\mu)$ is fixed.}
\label{tab:fit}     
\begin{center}
\begin{tabular}{llll}
\hline\noalign{\smallskip}
 & SPS1a & $\Delta_{masses}$ & $\Delta_{edges}$   \\
\noalign{\smallskip}\hline\noalign{\smallskip}
$m_0$ (GeV) & 100 & 3.9 & 1.2 \\
$m_{1/2}$ (GeV) & 250 & 1.7 & 1.0 \\
tan($\beta$) & 10 & 1.1 & 0.9 \\
$A_0$ (GeV) & -100 & 33 & 20 \\
\noalign{\smallskip}\hline
\end{tabular}
\end{center}
\vspace*{-0.5cm}  
\end{table}

\section{Spin Measurements}
If SUSY signals are observed at the LHC, it will be vital to measure the spins of 
the new particles to demonstrate that they are indeed the predicted super-partners. 
Two methods of measuring the spin are described below.

\subsection{Neutralino Spin Measurement}
\begin{figure}
\begin{center}
\includegraphics[width=0.4\textwidth,height=0.27\textwidth,angle=0]{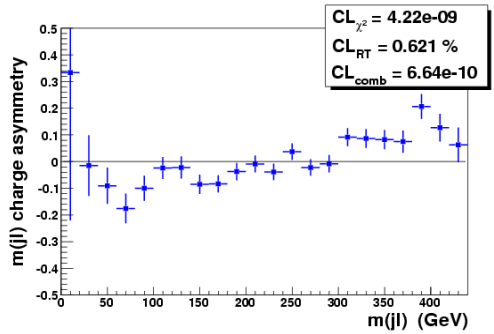}
\caption{The lepton charge asymmetry in the invariant mass of the lepton-jet for the SU3 point 
with an integrated luminosity of 30 fb$^{-1}$.}
\label{fig:spin_1}    
\end{center}
\end{figure}

The decay chain ${\widetilde q}_{L} \rightarrow q {\widetilde \chi}_{2}^{0} 
\rightarrow q {\widetilde l}_{R}^{\pm}l_{near}^{\mp} \rightarrow q {\widetilde \chi}_{1}^{0} 
l^{\pm}_{far} l^{\mp}_{near}$ provides a good oppurtunity to measure the spin of 
${\widetilde \chi}_{2}^{0}$ using the lepton charge asymmetry~\cite{Barr}. The squarks and 
sleptons are spin-0 particles and their decays are spherically symmetric, 
${\widetilde \chi}_{2}^{0}$ however has spin-1/2 and the angular distribution of its 
decay products is not spherically symmetric. This leads to a charge asymmetry of the 
invariant mass of the quark and the near lepton which is defined as:
\begin{equation}
A^{+-}=\frac{s^{+}-s^{-}}{s^{+}-s^{-}}\;\;\;\;\;\; s^{\pm}=\frac{d\sigma}{dm(ql^{\pm})}
\end{equation}
The asymmetry is suppressed by the fact that quark jets cannot be experimentally distinguished 
from anti-squark jets at the LHC, however at the LHC much more squark than antisquark will be 
produced, and therefore a residual asymmetry is still observable. This asymmetry is calculated 
for the SU3 point considering a fast simulation sample of 30 fb$^{-1}$ together with the most 
relevant SM backgrounds, $t\bar{t}$, W+jets, Z+jets. Events are selected by requiring two opposite 
sign leptons ($e$ or $\mu$) with $p_{T}>$10 GeV and $|\eta|<$2.5, 
a lepton isolation cut of $E_{T}<$10 GeV in a cone of 0.2, at least four jets with 
$E_{T}>$100, 50, 50, 50 GeV and $E_{T}^{\rm miss}>$100 GeV. The charge asymetry is plotted in 
Figure~\ref{fig:spin_1} by considering both the near and far leptons since these are not distinguishable 
for the SU3 point. The confidence level for the asymmetry distribution to be flat (spin-0) 
is calculated using two independent statistical methods as described in~\cite{Biglietti} and found to 
be $\sim10^{-9}$. It is observed that 10 fb$^{-1}$ of data would be sufficient to detect a non-zero 
charge asymmetry at the 99\% confidence level for the SU3 point.

\subsection{Slepton Spin Measurement} 
\begin{figure}
\begin{center}
\includegraphics[width=0.4\textwidth,height=0.28\textwidth,angle=0]{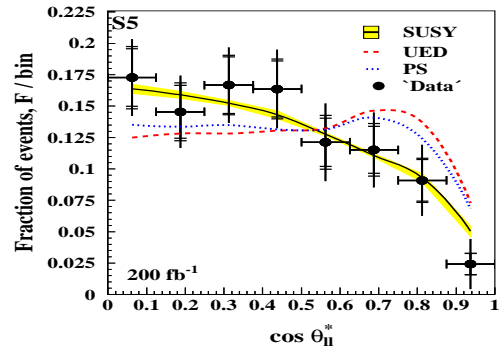}
\caption{$cos \theta_{ll}^{*}$ distribution for the Point 5 with an integrated 
luminosity of 200 fb$^{-1}$.}
\label{fig:spin_2}       
\end{center}
\end{figure}

The spin of the slepton can be measured using an angular variable which is 
sensitive to the polar angle in direct slepton pair production $q\bar{q} \rightarrow Z^{0}/\gamma 
\rightarrow {\widetilde l}^{+} {\widetilde l}^{-} \rightarrow {\widetilde \chi}_{1}^{0} 
l^{+} {\widetilde \chi}_{1}^{0}l^{-}$ as discussed in~\cite{Barr_slepton}. 
The angular variable is defined as 
$cos\theta_{ll}^{*} \equiv cos(2tan^{-1}e^{(\Delta \eta_{l^{+}l^{-}}/2)})= tanh \\
(\Delta \eta_{l^{+}l^{-}}/2)$
It is interpreted as the cosine of the polar angle between each lepton and the beam 
axis in the longitudinally boosted frame in which the pseudorapidities of the leptons 
are equal and opposite. This variable is on average smaller for SUSY than for the 
Universial Extra Dimensions (UED) so it can be employed as a spin-discriminant 
in slepton/Kaluza-Klein-lepton pair production in hadron colliders. 
The study was performed on Point 5 studied for the ATLAS TDR \\
~\cite{ATLASTDR} which has a phenomenology very similar to the one of point SU3.
A fast simulation sample of 200 fb$^{-1}$ is considered together with the major SM backgrounds. 
Events are selected by requiring two opposite sign and same flavor electrons or muons 
with $p_{T}(l_{1})>$40 GeV, $p_{T}(l_{2})>$30 GeV, $M_{ll}<$150 GeV, 
no jet with $E_{T}>$100 GeV, no b-jets, $E_{T}^{\rm miss}>$100 GeV, 
$M_{T2}<$100 GeV. The angular variable is plotted in Figure~\ref{fig:spin_2} 
together with the predictions for SUSY (black line), phase space (dotted blue line) and the 
UED (dashed red line). The yellow shaded band represent the SUSY expectation when all the SUSY 
particle masses are simultaneously changed by $\pm$20 GeV. As seen in the figure, the data 
points are much better matched to the slepton angular distribution than to either the 
phase-space one or the UED-like one. This shows that $cos \theta_{ll}^{*}$ does indeed measure 
the spin of the sleptons for this point. Further studies with the Snowmass points 
SPS1a, SPS1b, SPS3, SPS5~\cite{Allanach:2002nj} allow slepton spin determination with 100-300 
fb$^{-1}$ of data.

\section{Conclusions}
LHC brings experimental physics into a new territory. ATLAS can discover 
the SUSY particles up to 2-3 GeV mass scale if they exist. Understanding the detector 
response and the SM background will be a challenge at the beginning. Many techniques 
have been developed to measure the masses (edges, thresholds, mass differences) and 
spin of SUSY particles and to determine the underlying model parameters. 

\end{document}